\documentclass[11pt]{article}
\renewcommand{\footnoterule}{%
  \kern -3pt
  \hrule width \textwidth height 0.5pt
  \kern 2pt
}

\newcommand{\dfdt}{\frac{dF(T)}{dT}}

\newcommand{\sqrtflrw}{\sqrt{1-kr^{2}}}
\newcommand{\st}{\sin\theta}
\newcommand{\ct}{\cos\theta}
\newcommand{\sph}{\sin\varphi}
\newcommand{\cph}{\cos\varphi}

\usepackage{latexsym}
\usepackage{amssymb}
\usepackage{amsmath}
\usepackage{graphicx}
\usepackage{bm}
\usepackage{enumerate}
\usepackage{sectsty}
\usepackage{subfig}
\usepackage{float}
\usepackage{caption}
\usepackage{times,fancyhdr}
\usepackage{dsfont}
\usepackage{moresize}
\usepackage{scrextend}

\sectionfont{\normalsize}
\subsectionfont{\small}

\numberwithin{equation}{section}

\newcommand*{\Scale}[2][4]{\scalebox{#1}{$#2$}}%

\begin{document}

\pagestyle{fancy}
\fancyhead{} 
\fancyhead[OR]{\thepage}
\fancyhead[OC]{{\small{JUNCTION CONDITIONS FOR $F(T)$ GRAVITY}}}
\fancyfoot{} 
\renewcommand\headrulewidth{0.5pt}
\addtolength{\headheight}{2pt} 

\title{{\normalsize{\bf{JUNCTION CONDITIONS FOR F(T) GRAVITY FROM A VARIATIONAL PRINCIPLE}}}}
\author {{\small Jesse Velay-Vitow \footnote{jvelayvi@sfu.ca}} \\
\it{\small Department of Physics,  Simon Fraser University} \\
\it{\small Burnaby, British Columbia, V5A 1S6, Canada}
\\[-0.1cm]
\line(1,0){45}\\[0.1cm]
\and
{\small Andrew DeBenedictis \footnote{adebened@sfu.ca}} \\
\it{\small Department of Physics, Simon Fraser University}\\
\it{\small and}\\
\it{\small The Pacific Institute for the Mathematical Sciences,} \\
\it{\small Burnaby, British Columbia, V5A 1S6, Canada }
}
\date{{\small July 10, 2017}}
\maketitle
\begin{abstract}
\noindent We derive a general set of acceptable junction conditions for $F(T)$ gravity via the variational principle. The analysis is valid for both the traditional form of $F(T)$ gravity theory as well as the more recently introduced Lorentz covariant theory of Kr\v{s}\v{s}\'{a}k and Saridakis. We find that the general junction conditions derived, when applied to simple cases such as highly symmetric static or time dependent geometries (such as spherical symmetry) imply both the Synge junction conditions as well as the Israel-Sen-Lanczos-Darmois junction conditions of General Relativity. In more complicated scenarios the junction conditions derived do not generally imply the well-known junction conditions of General Relativity. However, the junctions conditions of de la Cruz-Dombriz, Dunsby, and S\'{a}ez-G\'{o}mez make up an interesting subset of this more general case.

\end{abstract}
\rule{\linewidth}{0.2mm}
\vspace{-1mm}
\noindent{\small PACS(2010): 04.20.Fy\;\; 11.10.Ef\;;\; MSC(2010): 58J32\;\;49S99}

\section{{Introduction}}\label{sec:intro}
Junction conditions play an important role in physical theories governed by differential equations. If a theory is governed by a set of differential equations in several domains, and one wishes to patch the solutions together at the junction between the domains, then one must know the acceptable continuity conditions to employ at the junction. These conditions are not arbitrary but must obey certain mathematical (and in the case of physical theories physically motivated) criteria. The importance of knowing the acceptable junction conditions of a theory is not just limited to sudden divides. The study of a theory's acceptable junction conditions also yields information regarding what must be continuous in continuous models. That is, even if the solution sought is continuous, knowledge of junctions conditions reveal at what level of derivative a discontinuity is allowed, and in which particular quantities  (for systems of several functions). 

One arena (out of many) where junction conditions are particularly important is that of gravitational field theory. A common application here is, for example, some system such as a star, where the surface of the star must be joined in some way to the outside vacuum domain. Even inside the star one might have several layers of different material which make up different domains in which the governing differential equations must be solved. Another example could be that of phase transitions in gravitational settings where the transition divide could be sudden or continuous. Currently, arguably the best theory of gravitation is the theory of general relativity and the junction conditions in general relativity have been studied in depth \cite{ref:grjcfirst} - \cite{ref:grjclast}. However, it is possible that gravitation is not described by general relativity, but instead by some other theory which yields general relativity in some limit. Each particular theory is accompanied with its own set of junction conditions which must be known if any patching of solutions is to be attempted. 

One particularly popular extension of general relativity is the class of theories which are nonlinear in the the Ricci scalar. These are the well-studied $F(R)$ theories. In these theories the acceptable junction conditions tend not to always coincide with those of general relativity \cite{ref:frjc1}, \cite{ref:frjc2}, \cite{ref:frjc3}. Another, less studied alternative to general relativity is $F(T)$ gravity. In $F(T)$ gravity the source of the gravitational field is not curvature, but instead torsion as described by the torsion tensor, $T^{\alpha}_{\;\;\beta\gamma}$, which is defined by the antisymmetry of the curvature-less Weitzenb\"{o}ck connection, $\underset{\mbox{\tiny{(W)}}}{\Gamma}^{\alpha}_{\;\;\gamma\beta}$ including the spin connection:
\begin{equation}
T^{\alpha}_{\;\;\beta\gamma}:=\underset{\mbox{\tiny{(W)}}}{\Gamma}^{\alpha}_{\;\;\gamma\beta}-\underset{\mbox{\tiny{(W)}}}{\Gamma}^{\alpha}_{\;\;\beta \gamma}=h_{a}^{\;\;\alpha}\left(\partial_{\beta}h^{a}_{\;\gamma}-\partial_{\gamma}h^{a}_{\;\beta}\right) + h_{a}^{\;\;\alpha} \omega^{a}_{\;\;b\beta}h^{b}_{\;\;\gamma} - h_{a}^{\;\;\alpha} \omega^{a}_{\;\;b\gamma}h^{b}_{\;\;\beta}\,. \label{eq:spintorsion}
\end{equation}
Here $h^{a}_{\;\mu}$ is the metric compatible tetrad and $\omega^{a}_{\;\;b\mu}$ is the inertial spin connection, which will be discussed shortly \footnote{Greek indices are spacetime indices whereas Latin indices are orthonormal Lorentz indices (sometimes also denoted by a hat).}. From the torsion tensor one constructs the torsion scalar, $T$:
\begin{equation}
 T:=\frac{1}{4}  T_{\alpha\beta\gamma}  T^{\alpha\beta\gamma} + \frac{1}{2}  T_{\alpha\beta\gamma}  T^{\gamma\beta\alpha}
     - T_{\alpha\beta}^{\;\;\;\alpha}T^{\gamma\beta}_{\;\;\;\;\gamma}\,, \label{eq:T}
\end{equation}
and the following action is formed:
\begin{equation}
I=\int \left\{\frac{F(T)}{2\kappa}+\mathcal{L}_{\mbox{\tiny{m}}}\right\}\,\mbox{det}[h^{a}_{\;\mu}]\,d^{4}x\,, \label{eq:gravaction}
\end{equation}
with $\kappa=8\pi$ and $\mathcal{L}_{\mbox{\tiny{m}}}$ the Lagrangian density for any matter fields which may be present.

In the traditional version of $F(T)$ gravity it is assumed that the inertial spin connection vanishes. Therefore, in that version of the theory one must choose a metric compatible tetrad which also yields zero for the components $\omega^{a}_{\;\;b\nu}$. If this is not done correctly, then one is inadvertently including inertial (non-gravitational) effects into the resulting equations of motion \cite{ref:kandp}, \cite{ref:kands}. This is essentially the origin of the ``non-Lorentz covariance'' of traditional $F(T)$ gravity \cite{ref:kands}, \cite{ref:noncov1}, \cite{ref:noncov2}, \cite{ref:cov}, \cite{ref:cov2} \cite{ref:newlorcovpaper}, and leads one to having to choose a ``good'' tetrad \cite{ref:goodbad}, \cite{ref:goodbadpres}. Finding such a tetrad is often non-trivial.

Progress has been made in $F(T)$ gravity in order to attempt to restore full Lorentz covariance. Kr\v{s}\v{s}\'{a}k and Saridakis have devised a straight-forward method to restore this covariance to the $F(T)$ field equations in \cite{ref:kands}. A more in-depth justification for the method has also been provided in \cite{ref:newlorcovpaper}. The method involves picking any metric compatible tetrad, and taking the $G\rightarrow 0$ limit of this tetrad. This $G=0$ ``reference tetrad'' is then used to calculate the torsion tensor (\ref{eq:spintorsion}). This torsion tensor, due to the absence of gravity in the $G\rightarrow 0$ limit, should vanish, and therefore one uses the condition $\lim_{G\rightarrow 0}T^{a}_{\;\beta\gamma}=0$ in order to solve for the components of the spin connection $\omega^{a}_{\;\;b\mu}$. It has also been shown in \cite{ref:kandp} that an appropriate spin connection can be viewed as a renormalization which yields a physically sensible stress-energy tensor (see also \cite{ref:griblpereira}). The resulting spin connection is then to be used in formulating the equations of motion.that which read:
 \begin{equation}
\Scale[0.95]{h^{-1} h^{a}_{\;\rho} \partial_{\mu} \Big( h \frac{F(T)}{dT} S_{a}^{\;\nu\mu} \Big)
- \frac{dF(T)}{dT} T_{\alpha\beta\rho} S^{\alpha\beta\nu}
+ \frac{1}{2} F(T) \delta_{\rho}^{\;\nu} + \frac{dF(T)}{dT} S_{a}^{\;\;\alpha\nu} h^{b}_{\;\;\rho}\omega^{a}_{\;\;b\alpha} = 8\pi \mathcal{T}_{\rho}^{\;\nu}}\,, \label{eq:eoms}
\end{equation}
with $\mathcal{T}_{\rho}^{\;\nu}$ the stress-energy tensor. The advantage of this version of the theory is that one does not have to look for a ``good'' tetrad but instead any metric compatible tetrad should suffice. The quantity $S_{\alpha\beta\gamma}$ is known as the superpotential (or modified torsion) and is defined as
\begin{equation} 
    S_{\alpha\beta\gamma} :=  K_{\beta\gamma\alpha}
      + g_{\alpha\beta} \,  T_{\sigma\gamma}^{\;\;\;\sigma}
      - g_{\alpha\gamma} \,  T_{\sigma\beta}^{\;\;\;\sigma}\,, \label{eq:stensor}
\end{equation}
with 
\begin{equation} 
    K_{\alpha\beta\gamma} :=
     \frac12 \left(  T_{\alpha\gamma\beta}
     +  T_{\beta\alpha\gamma} +  T_{\gamma\alpha\beta} \right)\, \label{eq:ktensor}
\end{equation}
the contorsion tensor.

The equations (\ref{eq:eoms}) are derived via variation of the action (\ref{eq:gravaction}) with respect to the tetrad, and in the special case where $F(T)=T$ one acquires the teleparallel equivalent of general relativity (TEGR), which is equivalent to Einstein theory save for a boundary term. In the case where $F(T)\neq T$ the resulting equations of motion, though more complicated, remain second-order, which differs from the corresponding $F(R)\neq R$ based theories. 

As mentioned earlier, this theory of gravity is not as well studied as curvature theories. However, it is a viable gravitational field theory as it yields general relativity results in the limit where terms in $F(T)$ which are nonlinear in $T$ may be neglected. Therefore it is useful to derive what the acceptable junction conditions are for this theory of gravity. One interesting set of junction conditions for the non covariant version of $F(T)$ gravity have been studied in \cite{ref:foftjc} from demanding the avoidance of thin-shells in the bulk equations of motion. In this paper we derive a set of junction conditions from the variational principle and deal in the covariant version of the theory, although the analysis also applies to the traditional version. The Lorentz covariant version of \cite{ref:kands} has not yet received as much attention as the traditional version. A study setting limits on possible non-linear torsion contributions in the Lorentz covariant theory may be found in \cite{ref:AandS}.

The $F(T)$ gravity has been used to study relativistic stars  \cite{ref:tamgood}, \cite{ref:abbas}, \cite{ref:abbas2}. Weyl static axially symmetric field equations in $F(T)$ gravity have been investigated in \cite{ref:nayeh}. In the cosmological framework $\Lambda$-CDM models have been studies in \cite{ref:Salako} in a thermodynamic perspective. Bianchi-I models have been analyzed in \cite{ref:sharifandrani}, \cite{ref:Paliathanasis}, whereas other anisotropic cosmological models have been studied in \cite{ref:rodriguesrahaman}. Studies of $F(T)$ cosmologies may also be found in the large volume \cite{ref:arXiv:1511.07586} (and references therein). A thorough analysis of general constraints from a cosmological perspective can be found in \cite{ref:cosconstr}. In the vein of metric affine theory, teleparallel gravity has been studied in \cite{ref:metaffine} as a particular case of generalized metric affine symmetry. A very thorough exposition on teleparallel gravity may be found in \cite{ref:teleparallelbook}.

\section{{Variational junction conditions}}\label{sec:jc}
\begin{figure}[!t]
\begin{center}
\includegraphics[scale=0.30, keepaspectratio=true]{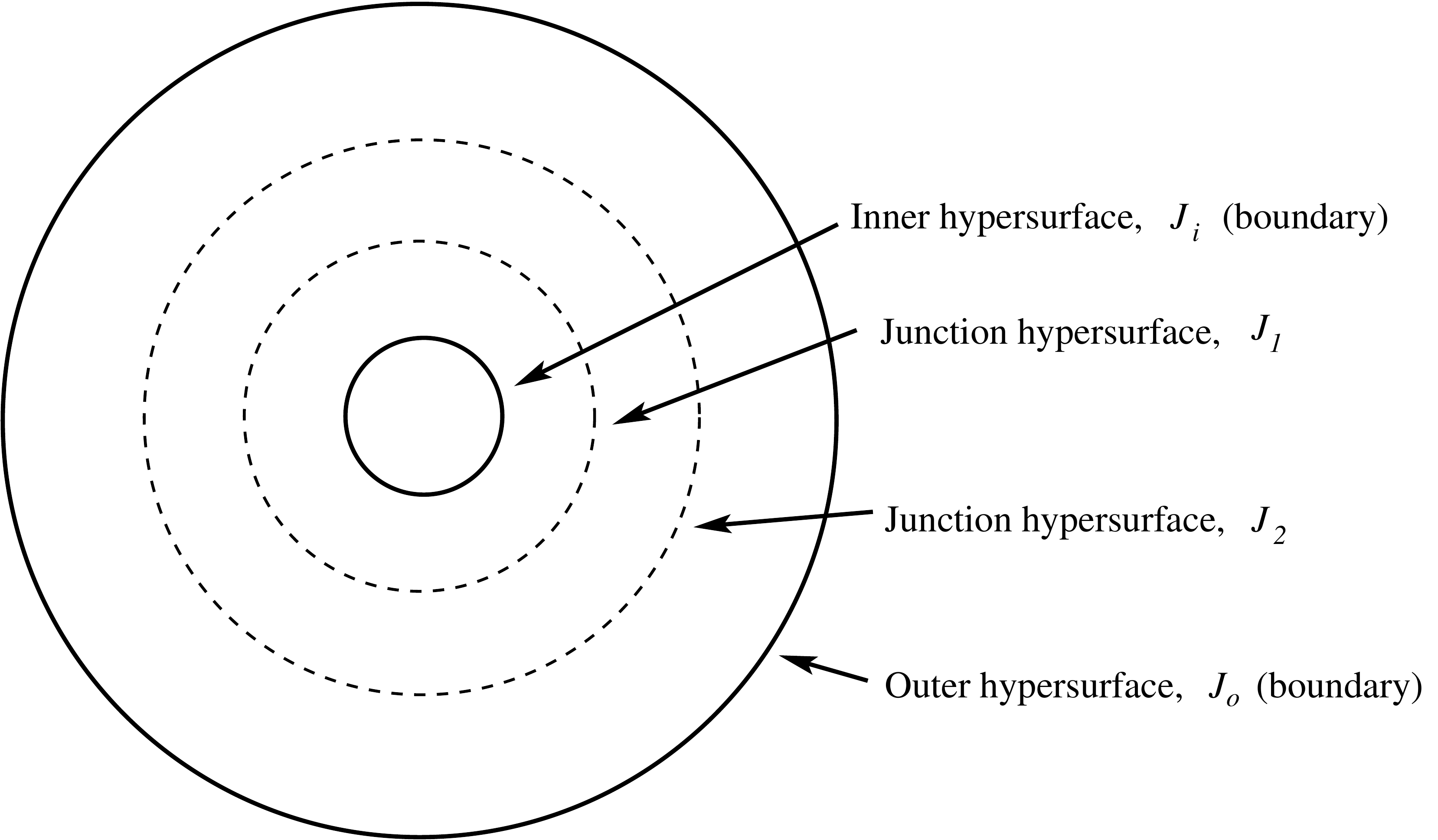}
\caption{{\small The domain of the differential equation of motion of interest divided into subdomains. The solid lines indicate the boundaries of the physical system ($J_{i}$ and $J_{o}$) and the dashed lines represent possible junctions ($J_{1}$ and $J_{2}$).}}
\label{fig:domains}
\end{center}
\end{figure}

One way to obtain a general set of junction conditions for a physical theory derived from an action is from the variational principle. We refer to these conditions as variationally admissible junction conditions. To illustrate the method we introduce it in a simplified one-dimensional context. Let us assume that we have some differential equation, which is derived via the variational principle of extremal action. Let the domain of this differential equation be the domain from the inner boundary surface ($J_{i}$ in figure \ref{fig:domains}) to the outer boundary surface ($J_{o}$ in figure \ref{fig:domains}). At both $J_{i}$ and $J_{o}$ some boundary conditions are employed. Perhaps Dirichlet conditions are employed there (eg. the field vanishes or is set to some other fixed asymptotic value), but the specific form is not relevant as long as some boundary conditions exist at these boundaries. There is some solution to this differential equation which obeys the boundary conditions. A physical example could be a star where we have a boundary condition at the center of the star ($J_{i}$) as well as a ``boundary'' condition far away from the star, $J_{o}$ (eg. Miknowski spacetime far away if $J_{o}$ is ``infinity'').

Now, one could also break up the problem into subdomains, as shown in the figure. This could be done, for example, because there are junction surfaces motivated by the particular problem one wants to solve (eg. stellar layers or a matter-vacuum boundary in the case of stars). The problem is still the same: One has a differential equation and this equation must be solved with specific boundary conditions at $J_{i}$ and $J_{o}$. Let us assume we have some action which depends on a dynamical field $\phi$ and its derivative, $\phi^{\prime}$ (first derivative only for simplicity):
\begin{equation}
I=\int\, \mathcal{L(\phi, \phi^{\prime})}\, dr\,, \label{eq:toyaction}
\end{equation}
where for the purpose of this toy model we label the integration variable $r$. Setting the variation of this action equal to zero yields:
\begin{equation}
\mbox{B.T.}|^{J_{o}}_{J_{i}} - \int_{J_{i+}}^{J_{o-}} \mbox{Bulk}\, dr = 0\,, \label{eq:variation}
\end{equation}
where B.T. refers to a ``boundary term'' and ``Bulk'' is the bulk integrand, which yields the equations of motion of the theory. Now, if one considers the same problem, but subdivides the interval due to the presence of junctions, as in figure \ref{fig:domains}, the variational principle of extremizing the action yields the following:
\begin{equation}
\Scale[0.95]{\underset{\epsilon\rightarrow 0}{\lim}\left\{\mbox{B.T.}|_{J_{i}}^{J_{1}-\epsilon} - \int_{J_{i+}}^{J_{1-}} \mbox{Bulk}\, dr + \mbox{B.T.}|_{J_{1}+\epsilon}^{J_{2}-\epsilon} - \int_{J_{1+}}^{J_{2-}} \mbox{Bulk}\, dr +
\mbox{B.T.}|_{J_{2}+\epsilon}^{J_{o}} - \int_{J_{2+}}^{J_{o-}} \mbox{Bulk}\, dr\right\} = 0}\,. \label{eq:juncvariations}
\end{equation}

For (\ref{eq:variation}) and (\ref{eq:juncvariations}) to be equal (note that the same problem is being solved) the boundary terms at the internal junctions ($J_{1}$ and $J_{2}$ here) need to be continuous as one approaches the junction from either side of the junction so that the boundary terms cancel out. This continuity across the junction hypersurfaces gives rise to the mathematically allowed variationally permitted junction conditions of the theory and yields a very general set of allowable junction conditions for a local theory \cite{ref:lanczosbook}. Physically, such continuity might imply the absence of infinitely thin shells which would otherwise be present due to the discontinuities. Note that even if one wishes to supplement the problem with further restrictions from, for example, some specific desired physics of a particular model, the continuity of the boundary terms would still need to hold. Hence those subsequent conditions need to be compatible with this continuity in order to have a well-posed variational problem.

In this section we will apply this analysis to the case of Lorentz covariant $F(T)$ gravity but we first note a few issues: Firstly, the variationally admissible junction conditions are derived from an action, but the action is not unique. One may always add a total divergence to the Lagrangian without affecting the bulk equations of motion. This addition will modify the surface term, and hence modify the junction conditions. We assume here that the action is minimal, meaning that the gravitational Lagrangian is of the form in (\ref{eq:gravaction}) and possesses no extra expressions which could potentially affect the resulting surface terms. Secondly, although it is possible that in some very special cases the surface terms potentially arising from the matter Lagrangian could cancel out discontinuities in the gravitational surface term (and vice-versa), in the generic setting this is generally not the case. We therefore do not consider this possibility and assume that the gravitational and matter surface terms (if present) are each individually continuous. Also, the mixing of gravitational field and matter in a non-minimal coupling theory (such as \cite{ref:lobononmin}, \cite{ref:DandS}, \cite{ref:DandS2}) would potentially yield a different set of junction conditions than those derived here. We consider minimally coupled models only.

Before proceeding to the $F(T)$ case it is perhaps interesting to first briefly look at general relativity in this context. Variation of the Einstein-Hilbert action yields the following boundary result:
\begin{equation}
  \left(\delta I\right)_{|\partial D} = - \frac{1}{16\pi} \int_{\partial D} \delta \left[g^{\mu\nu} \Gamma^{\lambda}_{\;\lambda \mu} - g^{\lambda\mu} \Gamma^{\nu}_{\;\lambda\mu}\right]\hat{n}_{\nu}\, d^{3}v \label{eq:einstvariation}
\end{equation}
and continuity of this boundary term would represent the variationally admissible junction conditions for general relativity at junction hypersurfaces. Note however that if on the inner and outer boundaries ($J_{i}$ and $J_{o}$ of figure (\ref{fig:domains})) one imposes Dirichlet boundary conditions, $\delta (g_{\alpha\beta})_{|\partial D}=0$, this is not necessarily sufficient to make the boundary terms vanish there. In that case one still has surviving quantities which may be written as \cite{ref:dasdebbook}
\begin{equation}
 \left(\delta I\right)_{|\partial D}= -\frac{1}{8\pi}\int_{\partial D} \delta K\,d^{3}v \,, \label{eq:einstDbound}
\end{equation}
for the surface term, which may be non-zero ($K$ being the trace of the extrinsic curvature). This is the origin of the famous Gibbons-Hawking surface term of general relativity which must be added to the action in order to have a well-posed variational problem \cite{ref:gibhawk}. (In $F(T)$ gravity such a term analogous to (\ref{eq:einstDbound}) does not arise, since as we'll see below the Dirichlet condition is sufficient to render the surface term to zero.) The appearance of such terms is discussed in detail in \cite{ref:joint}. The effect of other boundary terms in the analog of the torsion equivalent of Gauss-Bonnet cosmology has been studied in \cite{ref:alvarorecent}. Regarding junctions, and therefore relevant to the the study here, the Israel-Sen-Lanczos-Darmois (ISLD) conditions of general relativity \cite{ref:I} - \cite{ref:D} comprise a sufficient set of junction conditions to make the boundary term of (\ref{eq:einstvariation}) (or (\ref{eq:einstDbound})) continuous at a junction, supplemented with the (usual) condition that the first fundamental form must also be continuous \cite{ref:nspg}. The junctions conditions of the various authors ISLD are not exactly equivalent to each other but in the literature the continuity of the second fundamental form often falls under the umbrella of the ISLD conditions. It may also be interesting to note that general relativity and the teleparallel equivalent of general relativity may possess different junction conditions even though they have the same equations of motion, since the Lagrangians differ by a total divergence:
\begin{equation}
 R=-T-2\nabla^{\alpha}\left(T^{\beta}_{\;\alpha\beta}\right)\,, \label{eq:TRreln}
\end{equation}
which translates into differing surface terms for the two theories. This is an example of a situation where analyzing junctions from the equations of motion is not sufficient to determine the full junction conditions of a theory. We next proceed with studying the junction conditions for $F(T)$ gravity.

\subsection{Junction conditions for $F(T)$ gravity}
To determine the variationally admissible junction conditions for $F(T)$ gravity one needs to vary the action (\ref{eq:gravaction}) with respect to the tetrad but the surface terms must not be discarded. The resulting bulk integrand implies the equations of motion (\ref{eq:eoms}) and we find that the resulting surface term may be written as:
\begin{equation}
\left(\delta I\right)_{|\partial D}= \frac{1}{2\kappa}\int_{\partial D} \frac{d\,F(T)}{dT} S_{a}^{\;\rho\sigma} \delta h^{a}_{\;\sigma}\, h\,  \hat{n}_{\rho}\, d^{3}\Sigma\,, \label{eq:bt}
\end{equation}
where $\partial D$ indicates the boundary of the domain, $\hat{n}_{\rho}$ denote the components of the unit normalized outward pointing normal (co)vector of the junction surface, and $h$ the determinant of the tetrad. It is an extremely long calculation to confirm that no other surface terms arise from the variation of the action via, for example, some hidden four-divergence when the action is explicitly written in terms of the tetrad and its derivatives, but we have done so. We make the (mild) assumption that the manifold is orientable in the sense that a unique definition of ``outward'' exists. Note that the surface terms arising from the spin connection are present in $T$ and the superpotential. We should also note that we consider here the scenarios where the spin connection is considered non-dynamical and fixed. This is compatible with both the traditional $F(T)$ gravity theory (where $\omega^{ab}_{\;\;\;\mu}=0$ is assumed from the start, and hence non-dynamical) as well as the version of the covariant theory of \cite{ref:kands} where the spin connection is seen as a function of a non-dynamical reference tetrad, yielding a bitetrad theory \cite{ref:newlorcovpaper}. In teleparallel theories one may wish to consider the spin connection also as a dynamical variable and therefore it is possible to also consider variations of the spin connection to yield various covarinatizatons of the theory \cite{ref:cov}. In these latter theories it is possible that extra surface terms arise as well as (2.7). If those surface terms do not vanish on the junctions then one may need to demand continuity of those terms as well. The general variationally admissible junction condition of $F(T)$ gravity in the current case is essentially the requirement that (\ref{eq:bt}) be continuous across the junction hypersurface. However, this is rather general and not very enlightening for practical calculations so below we shall analyze some specific scenarios of physical interest. Before proceeding to these, we make the following comments regarding the general case:\\[0.2cm]

\begin{itemize}
 \item A sufficient condition for the continuity of (\ref{eq:bt}) is that the integrand is continuous on $\partial D$. We will restrict subsequent analysis to this case.
 \item We will generally assume that the tetrad is continuous on $\partial D$. This also implies a common coordinate system on either side of the junction. This is analogous to demanding that the metric is continuous in general relativity. This ensures that $h$ and  $\delta h^{a}_{\;\sigma}$ are continuous across $\partial D$. 
 \item The Dirichlet conditions $\delta h^{a}_{\;\sigma}=0$, though sufficient to make (\ref{eq:bt}) vanish and hence yield continuity at a junction, can generally not be set as these conditions may be incompatible with the boundary conditions imposed on the problem at $J_{i}$ and $J_{o}$.
 \item In the case of TEGR, $dF(T)/dT=1$, and therefore, subject to the above conditions listed, the junction conditions of TEGR boil down to only requiring continuity of the superpotential in the normal direction, $[S_{a}^{\;\rho\sigma} \hat{n}_{\rho}]_{\pm}=0$.
 \item For $F(T)\neq T$, the quantity $dF(T)/dT$ is a function of $T$ and therefore continuity of the torsion scalar is sufficient to make this quantity continuous.
 \item Since $T=T^{a}_{\;\mu\nu}S_{a}^{\;\mu\nu}$ (with $S_{a}^{\;\mu\nu}$ being constructed out of the torsion tensor as in (\ref{eq:stensor}) and (\ref{eq:ktensor})), under the above assumptions it is possible to satisfy junction conditions simply by demanding that \emph{all} components of the torsion tensor be continuous. However, this is often too restrictive a condition. It is analogous to demanding that the Riemann tensor must be continuous in general relativity. This imposition will certainly satisfy the junction conditions of general relativity, but it is too severe a condition to describe, for example, a matter-vacuum junction where the energy density $(T^{0}_{\;\,0} \propto G^{0}_{\;\,0} =(1/4)\, \varepsilon^{\alpha 0 \mu\nu} R_{\mu\nu}^{\;\:\:\:\,\rho\sigma}\varepsilon_{\rho\sigma 0 \alpha})$ drops to zero discontinuously. 
\end{itemize}

\subsubsection{CASE I: Static spherical symmetry}
Here we study what may arguably be the most physically relevant special case, that of static spherical symmetry. This is relevant, for example, in theoretical models of calm stars, idealized galaxies, or any distribution of matter which may be approximated as spherical and time independent. We take the following tetrad
\begin{equation}
\left[h^{a}_{\;\mu}\right]=\left( \begin{array}{cccc}
A(r) & 0 & 0 & 0\\
0 & B(r) & 0 & 0 \\
0 & 0 & r & 0 \\
0 & 0 & 0 & r\sin\theta  \end{array} \right)\,, \label{eq:sssdiagtet}
\end{equation}
which is compatible with the well-known following metric:
\begin{equation}
 ds^{2}=A^{2}(r) dt^{2}-B^{2}(r) dr^{2}-r^{2}d\theta^{2}-r^{2}\sin^{2}\theta\,d\varphi^{2}\,. \label{eq:sphereline}
\end{equation}
Since we are employing the Lorentz covariant version of the theory of \cite{ref:kands}, we are free to choose the diagonal tetrad (\ref{eq:sssdiagtet}). The non-zero inertial spin connection components for this case are as follows:
\begin{equation}
\omega^{\hat{{r}}\hat{{\theta}}}{}_\theta = - \omega^{\hat{\theta}\hat{r}}{}_{\theta} = 1, \quad
\omega^{\hat{r}\hat{\varphi}}{}_{\varphi} = - \omega^{\hat{\varphi}\hat{r}}{}_{\varphi} = \sin\theta, \quad
\omega^{\hat{\theta}\hat{\varphi}}{}_{\varphi} = - \omega^{\hat{\varphi}\hat{\theta}}{}_{\varphi} = \cos\theta
\,. \label{eq:sssspincon}
\end{equation}

For static spherical symmetry the junctions of interest reside on $r=\mbox{const.}$ surfaces. Therefore, the unit outward normal points in the radial direction:
\begin{equation}
 \hat{n}_{\rho}=\frac{\delta^{\hat{r}}_{\;\rho}}{\sqrt{|\delta^{\hat{r}}_{\;\alpha}\,h_{a}^{\;\alpha}\,h^{a\beta}\,\delta^{\hat{r}}_{\;\beta}|}} =\delta^{\hat{r}}_{\;\rho} \left|B(r)\right|\,. \label{eq:sssnorm}
\end{equation}
Note that from continuity of the tetrad, the unit normal is already continuous at the junction hypersurface. Demanding the continuity of the integrand in (\ref{eq:bt}), in this case, yields the following set of conditions:
\begin{equation}
  \left[S_{a}^{\;\mathrm{r}\sigma}\right]_{\pm}=0 \quad \mbox{and} \quad \left[T\right]_{\pm}=0\,, \label{eq:ssscontconds}
\end{equation}
where the index $\mathrm{r}$ indicates that this index is fixed to the $\mathrm{r}$ component. (Recall that the unit normal is already continuous due to continuity of the tetrad, and that $dF(T)/dT$ can be made continuous by demanding continuity of $T$, yielding the second condition in (\ref{eq:ssscontconds}).) We find for the relevant non-zero components of the superpotential:
\begin{subequations}
\begin{align}
S_{\hat{\theta}}^{\;\;r \theta}=& \frac{A(r)+A_{,r}(r)r-A(r)B(r)}{A(r)B^{2}(r)r^{2}}\,,\label{eq:sssSa}\\[0.1cm] 
S_{\hat{\varphi}}^{\;\;r \varphi}=&\frac{A(r)+A_{,r}(r)r-A(r)B(r)}{A(r)B^{2}(r) r^{2}\sin(\theta)}\,,\label{eq:sssSb}\\[0.1cm]
S_{\hat{t}}^{\;\;r t}=&2\frac{1-B(r)}{A(r)B^{2}(r)r}\,,\label{eq:sssSc}
\end{align}
\end{subequations}
where the comma in the subscript denotes partial differentiation. We also have, for the torsion scalar:
\begin{equation}
T=\frac{2\left(B(r)-1\right)\left(A(r)B(r)-2A_{,r}(r)r-A(r)\right)}{A(r)B^{2}(r)r^{2}}\,. \label{eq:sssT}
\end{equation}
Note that with the assumption that the continuity of $A(r)$ and $B(r)$ already hold (via continuity of the tetrad), these continuity conditions on $S_a^{\;\;\mathrm{r}\nu}$ and $T$ imply the residual condition that the derivative 
\begin{equation}\label{eq:sssresidcond}
\left[A_{,r}(r)\right]_{\pm}=0
\end{equation}
must be continuous at the junction surface. It is interesting that this turns out to be the same condition implied by both the Synge conditions and the ISLD conditions in general relativity. The Synge conditions of general relativity read \cite{ref:syngebook}:
\begin{equation}
 \left[\mathcal{T}^{\mu}_{\;\:\nu}\hat{n}_{\mu}\right]_{\pm}=0\,. \label{eq:syncond}
\end{equation}
For the normal given in (\ref{eq:sssnorm}) this requires that the radial pressure, and by extension the $G_{r}^{\;r}=8\pi\mathcal{T}_{r}^{\;r}$ Einstein field equation, must be continuous at the junction. Let us now look at the radial pressure in this static spherically symmetric case for $F(T)$ gravity. The relevant $F(T)$ field equation is (suppressing $r$ dependence of the functions)

\begin{equation}
\frac{ \left[ 4r \left( B -2 \right) A_{,r} +4A  (B-1) \right] \dfdt + F(T){r}^{2}B^{2}A}{2{r}^{2}AB^{2}}=4\pi\mathcal{T}_{r}^{\;r} \, .\label{eq:ssspres}
\end{equation}
It can be seen that, since $A(r)$ and $B(r)$ are continuous, the residual condition (\ref{eq:sssresidcond}) is sufficient to satisfy continuity of (\ref{eq:ssspres}). Further, the ISLD junction conditions generally read
\begin{equation}
 \left[K_{\mu\nu}\right]_{\pm}=0\,, \label{eq:isldcond}
\end{equation}
with $K_{\mu\nu}$ the extended extrinsic curvature of the junction hypersurface. (We use the four-dimensional extended extrinsic curvature here, with one eigenvalue zero, but could easily have used its 3-dimensional counterpart in lieu. Also, we refer to the extrinsic curvature with the Christoffel connection, since we wish to make comparisons with the results of general relativity.) For an $r=\mbox{const.}$ hypersurface, the extrinsic curvature is calculated as:
\begin{equation}
 K_{\mu\nu}=\left[\begin{array}{cccc}
-\frac{A_{,r}(r)}{2\sqrt{B(r)}} & 0 & 0 & 0 \\
0 & 0 & 0 & 0\\
0 & 0 & \frac{r}{\sqrt{B(r)}} & 0\\
0 & 0 & 0 & \frac{r\sin^{2}(\theta)}{\sqrt{B(r)}} 
\end{array} \right]_{|r=r_{0}} \,. \label{eq:sssISLD}
\end{equation}
Again noting the continuity of $A(r)$ and $B(r)$, the condition (\ref{eq:sssresidcond}) implies continuity of (\ref{eq:sssISLD}) and hence a similar result as in general relativity holds for static spherical symmetry; namely, continuity of the radial pressure or continuity of the second fundamental form at the junction hypersurface. One can confirm that the same conclusions can be drawn if one works in the non-Lorentz covariant version of the theory but with the properly rotated tetrad of \cite{ref:goodbadpres}, \cite{ref:tamgood} yielding zero inertial spin connection.

\subsubsection{CASE II: Homogeneous spherically symmetric black hole interiors}
Here we look at the time dependent analog of the static spherically symmetric scenario considered previously. This is useful for the study of certain black hole interiors (sometimes known as ``T-spheres'' \cite{ref:ruban1}-\cite{ref:tlast}) and certain cosmological models. Black hole interiors in $F(T)$ gravity have been studied in \cite{ref:aftergood}. The line-element for these metrics takes on the following form \footnote{As an example, the Schwarzschild interior in this chart would correspond to the line element

$$ds^{2}_{\mbox{\tiny{Schw}}}= \frac{dt^{2}}{\frac{2M}{t}-1} - \left(\frac{2M}{t}-1\right) dy^{2} -t^{2}\,d\theta^{2} -t^{2}\sin^{2}\theta\,d\varphi^{2}\,,$$ with $0 < t < 2M$.}
\begin{equation}
 ds^{2}=A^{2}(t) dt^{2}-B^{2}(t) dy^{2}-t^{2}d\theta^{2}-t^{2}\sin^{2}\theta\,d\varphi^{2}\,, \label{eq:tsphereline}
\end{equation}
and the corresponding tetrad chosen is
\begin{equation}
\left[h^{a}_{\;\mu}\right]=\left( \begin{array}{cccc}
A(t) & 0 & 0 & 0\\
0 & B(t) & 0 & 0 \\
0 & 0 & t & 0 \\
0 & 0 & 0 & t\sin\theta  \end{array} \right)\,. \label{eq:tdiagtet}
\end{equation}
The results here are rather similar to the static spherically symmetric case so we only summarize briefly. The resulting inertial spin connection is given by \footnote{In the $G\rightarrow 0$ limit it may be the case that the $t$ coordinate becomes spacelike and the $y$ coordinate time-like, but we retain the labels $t$ and $y$ for notational compatibility with the rest of the calculations in this section.}:
\begin{equation}
\omega^{\hat{t}\hat{\theta}}{}_\theta = - \omega^{\hat{\theta}\hat{t}}{}_{\theta} = -1, \quad
\omega^{\hat{t}\hat{\varphi}}{}_{\varphi} = - \omega^{\hat{\varphi}\hat{t}}{}_{\varphi} = -\sin\theta, \quad
\omega^{\hat{\theta}\hat{\varphi}}{}_{\varphi} = - \omega^{\hat{\varphi}\hat{\theta}}{}_{\varphi} = \cos\theta
\,. \label{eq:tspincon}
\end{equation}
The junctions of interest in ``T-domains'' correspond to $t=\mbox{const.}$ hypersurfaces so that the unit normal points in the $t$ direction and the relevant components of the superpotential are:
\begin{subequations}
\begin{align}
S_{\hat{\theta}}^{\;\;t \theta}=& \frac{A(t)B(t)-B_{,t}(t)t-B(t)}{A^{2}(t)B(t)t^{2}}\,,\label{eq:tSa}\\[0.1cm] 
S_{\hat{\varphi}}^{\;\;t \varphi}=&\frac{A(t)B(t)-B_{,t}(t)t-B(t)}{A^{2}(t)B(t)t^{2}\sin(\theta)}\,,\label{eq:tSb}\\[0.1cm]
S_{\hat{y}}^{\;\;t y}=&2\frac{A(t)-1}{A^{2}(t)B(t)t}\,,\label{eq:tSc}
\end{align}
\end{subequations}
and the torsion scalar is given by
\begin{equation}
T=\frac{2\left(1-A(t)\right)\left(A(t)B(t)-2B_{,t}(t)t-B(t)\right)}{A^{2}(t)B(t)t^{2}}\,. \label{eq:tT}
\end{equation}
As before, noting that continuity of the tetrad holds, demanding continuity of (\ref{eq:tSa}-c) and of (\ref{eq:tT}) yields the residual condition that the following first derivative must be continuous:
\begin{equation}\label{eq:tresidcond}
\left[B_{,t}(t)\right]_{\pm}=0\,.
\end{equation}
The Synge condition (\ref{eq:syncond}) in this case implies that the $t-t$ equation of motion must be continuous and we wish to check if this condition holds in the $F(T)$ case. The relevant equation of motion here is
\begin{equation}
-\frac{ \left[ 4t \left( A -2 \right) B_{,t} +4B  (A-1)  \right] \dfdt -F(T) {t}^{2}A^{2}B}{2{t}^{2}A^{2}B}=4\pi\mathcal{T}_{t}^{\;t} \,.\label{eq:tedens}
\end{equation}
and the extended extrinsic curvature is
\begin{equation}
 K_{\mu\nu}=\left[\begin{array}{cccc}
0 & 0 & 0 & 0\\
0 & -\frac{B_{,t}(t)}{2\sqrt{A(t)}} & 0 & 0 \\
0 & 0 & -\frac{t}{\sqrt{A(t)}} & 0\\
0 & 0 & 0 & -\frac{t\sin^{2}(\theta)}{\sqrt{A(t)}} 
\end{array} \right]_{|t=t_{0}} \,. \label{eq:tISLD}
\end{equation}
Note that (\ref{eq:tresidcond}) implies both the continuity of (\ref{eq:tedens}) as well as (\ref{eq:tISLD}) and hence again in this case the $F(T)$ junction conditions turn out to be similar to the Synge and ISLD conditions of general relativity. A similar conclusion can be drawn utilizing the non Lorentz covariant theory without spin connection and the T-domain rotated tetrad of \cite{ref:aftergood}.

\subsubsection{CASE III: FLRW cosmology}
We now turn our attention to the interesting arena of Friedmann-Lema\^{i}tre-Robertson-Walker cosmology. The line element of interest is of the following well-known form:
\begin{equation}
 ds^{2}=dt^{2} - a^{2}(t)\left[\frac{dr^{2}}{1-kr^{2}}+r^{2}d\theta^{2}+r^{2}\sin^{2}\theta d\varphi^{2}\right]\,, \label{eq:flrwline}
\end{equation}
admitting a tetrad of 
\begin{equation}
\left[h^{a}_{\;\mu}\right]=\left( \begin{array}{cccc}
1 & 0 & 0 & 0\\
0 & \frac{a(t)}{\sqrt{1-kr^{2}}} & 0 & 0 \\
0 & 0 & a(t)r & 0 \\
0 & 0 & 0 & a(t)r\sin\theta  \end{array} \right)\,. \label{eq:flrwdiagtet}
\end{equation}
The inertial spin connection components in this case are
\begin{equation}
\omega^{\hat{r}\hat{\theta}}{}_\theta = - \omega^{\hat{\theta}\hat{r}}{}_{\theta} = 1, \quad
\omega^{\hat{r}\hat{\varphi}}{}_{\varphi} = - \omega^{\hat{\varphi}\hat{r}}{}_{\varphi} = \sin\theta, \quad
\omega^{\hat{\theta}\hat{\varphi}}{}_{\varphi} = - \omega^{\hat{\varphi}\hat{\theta}}{}_{\varphi} = \cos\theta
\,, \label{eq:flrwspincon}
\end{equation}
(in the absence of gravity $k$ must be set to zero and we set $a(t)=1$). The torsion scalar is
\begin{equation}
 T=- \frac{2k+(4/r^{2})(\sqrt{1-kr^{2}}-1) +6 a^{2}_{,t}(t)}{a^{2}(t)}\,. \label{eq:flrwT}
\end{equation}
For a time-like unit vector adapted to the time coordinate of (\ref{eq:flrwline}), the relevant superpotential components are:
\begin{subequations}
\begin{align}
S_{\hat{r}}^{\;\;t r}=& -2\sqrt{1-kr^2}\,\frac{a_{,t}(t)}{a^{2}(t)}\,,\label{eq:flrwSa}\\[0.1cm] 
S_{\hat{\theta}}^{\;\;t \theta}=&-\frac{2}{r}\,\frac{a_{,t}(t)}{a^{2}(t)}\,,\label{eq:flrwSb}\\[0.1cm]
S_{\hat{\varphi}}^{\;\;t \varphi}=&-\frac{2}{r \sin\theta}\,\frac{a_{,t}(t)}{a^{2}(t)}\,.\label{eq:flrwSc}
\end{align}
\end{subequations}
With continuous tetrad (\ref{eq:flrwdiagtet}), demanding continuity of (\ref{eq:flrwT}) and (\ref{eq:flrwSa}-c) requires the residual condition
\begin{equation}
 \left[a_{,t}(t)\right]_{\pm}=0\, \label{eq:flrwresid}
\end{equation}
at the junction hypersurface. One could have perhaps guessed this condition in this case as it is the only nontrivial first derivative for FLRW cosmology. Since every non-zero component of the extrinsic curvature on a $t=\mbox{const.}$ surface of the FLRW spacetime involves the metric functions and the first derivative of $a(t)$, the ISLD conditions are again implied by the $F(T)$ junctions conditions.

One may also utilize a rotated tetrad and the non-covariant theory, without the spin-connection. This tetrad is given by \cite{ref:goodbadpres}
\begin{equation}
\left[h^{a}_{\;\mu}\right]=\Scale[0.70]{\left( \begin{array}{cccc}
1 & 0 & 0 & 0\\
0 & a(t)\cph\st/\sqrt{1-kr^{2}} & a(t)\st\sph/\sqrtflrw & {a(t)\ct/\sqrtflrw} \\
0 & a(t)r\left(\sqrtflrw r\ct\cph-r\sph\right) & a(t)r\left(r\cph+\sqrtflrw\ct\sph\right) & -a(t)r\sqrtflrw\st \\
0 & -a(t)r\st\left(r\ct\cph-\sqrtflrw\sph\right) & a(t)r\st\left(\sqrtflrw\cph-r\ct\sph\right) & -a(t)r^{2}\sin^{2}\theta  \end{array} \right)}\,. \label{eq:flrwrottet}
\end{equation}
The superpotential components are rather complicated in this case but it can be verified that the only derivative present in $S_{a}^{\;\;t\sigma}$ is the first derivative of $a(t)$ as is true for the torsion scalar. Hence the same result as above is implied with this tetrad.

\subsubsection{More general cases}
Assuming a continuous tetrad, the general junction condition derived here requires continuity of $T$ as well as $S_{a}^{\;\:\rho\sigma}\hat{n}_{\rho}$. If these conditions are met then one has a well-posed variationally motivated physical solution. In the special cases above it was shown that these junction conditions imply the well-known junction conditions of Synge as well as those of Israel, Sen, Lanczos and Darmois that arise in general relativity. In a more general setting the Synge and ISLD conditions in general relativity are usually not equivalent. It is of interest to see what the $F(T)$ junction conditions imply with respect to the general relativity conditions in a more generic setting, such as relaxing the spherically symmetric tetrad of (\ref{eq:sssdiagtet}) to one where the functions $A$ and $B$ depend on both radius and time\footnote{The $G\rightarrow 0$ limit of this case is the same as in the static scenario studied previously and hence yields a similar spin connection as (\ref{eq:sssspincon}).}. It is natural in this case to consider junctions whose hypersurface reside on the level curve $r=h(t)$ (the problem is still of codimension 1). This is applicable, for example, in the scenario of collapsing stars. Some of the expressions are slightly lengthy so for this case we summarize the following observations:
\begin{itemize}
 \item The torsion scalar $T$ contains the derivative $A_{,r}(r,t)_{r=h(t)}$ and hence this derivative must be continuous. (In fact, $T$ has the same form as (\ref{eq:sssT}), save for the extra time dependence in the functions.)
 \item The relevant quantities of $S_{a}^{\;\:\rho\sigma}\hat{n}_{\rho}$ (specifically the components $S_{a}^{\;\:t\sigma}$ and $S_{a}^{\;\:r\sigma}$) contain both $A_{,r}(r,t)_{r=h(t)}$ as well as $B_{,t}(r,t)_{r=h(t)}$. Therefore $B_{,t}(r,t)_{r=h(t)}$ is also required to be continuous.
 \item To compare with the ISLD conditions of general relativity, we note that the extrinsic curvature of an $r=h(t)$ hypersurface possesses $r$ and $t$ derivatives of \emph{both} $A(r,t)$ and $B(r,t)$. Therefore the ISLD conditions are not implied.
 \item To compare with the Synge conditions of general relativity we note that the $t-t$ field equation and the $r-r$ field equation contain $A_{,r}(r,t)$ and $B_{,r}(r,t)$ and the $t-r$ and $r-t$ field equations contain derivatives with respect to $t$ and $r$ (or mixed), up to second order. Therefore the Synge condition is not implied.
\end{itemize}
We note here a peculiarity in this case: Even in the Lorentz covariant version of the theory as calculated here, the equations of motion seem to contain a pathology for the case when both $r$ and $t$ dependence is present in the tetrad functions $A$ and $B$, in that they imply that $\mathcal{T}_{tr}\neq\mathcal{T}_{rt}$ when $F(T)\neq T$. This is reminiscent of the issue in non-Lorentz covariant $F(T)$ gravity when one chooses a ``bad'' tetrad \cite{ref:goodbad}. As we are not aware of any ``good'' tetrad in the literature for the case of $A(r,t)$ and $B(r,t)$ we cannot comment further and this requires further study. We should mention though that this has no affect on the form of the general junction condition (\ref{eq:bt}) whose continuity remains a valid set of conditions for $F(T)$ gravity.

In even more general cases one cannot say that the $F(T)$ junctions conditions, requiring continuity of $T$ and $S_{a}^{\;\:\rho\sigma}\hat{n}_{\rho}$, yield the same junction conditions as in general relativity. This, of course, is to be expected since $F(T)$ gravity theory is not general relativity, and even TEGR differs from Einstein theory by boundary terms. 

\section{Concluding remarks}
By demanding uniqueness of the variational problem we derived a set of general junction conditions for $F(T)$ gravity (traditional as well as the covariant version of \cite{ref:kands}) for the problem of codimension 1. These junction conditions require continuity of the integral (\ref{eq:bt}). Under mild assumptions, a very general set of conditions which can enforce this continuity are continuity of the torsion scalar, and continuity of $S_{a}^{\;\:\rho\sigma}\hat{n}_{\rho}$, where $\hat{n}$ is the unit (outward) pointing normal to the junction hypersurface. It is interesting that in certain cases these conditions for $F(T)$ gravity are analogous to the Synge and the Israel-Sen-Lanczos-Darmois conditions of general relativity. Several physically important special cases were studied where this equivalence holds or partially holds. In general this equality is not the case but the results of \cite{ref:foftjc}, where $F(T)$ junction conditions were studied for the first time from demanding the avoidance of bulk thin shells in the non Lorentz covariant version of the theory, make up an interesting subset of conditions which may satisfy continuity of (\ref{eq:bt}).

\section*{Acknowledgments}
We are grateful to S. Iliji\'{c} of the University of Zagreb for helpful discussions regarding this project and for careful checking of some of the calculations.



\linespread{0.6}
\bibliographystyle{unsrt}

\end{document}